\documentclass[11pt,twoside]{article}

\usepackage{asp2014}

\aspSuppressVolSlug
\resetcounters

\begin{document}

\title{Supporting FAIR Principles in the Astrophysics Community:\\the European Experience}

\author{Marco~Molinaro,$^1$ Mark~Allen,$^2$ Fran\c{c}ois~Bonnarel,$^2$ Fran\c{c}oise~Genova,$^2$ Markus Demleitner,$^3$ Kay~Graf,$^4$  Dave~Morris,$^5$ Enrique~Solano,$^6$ and Andr\'{e}~Schaaff$^2$}
\affil{$^1$INAF - Osservatorio Astronomico di Trieste, Trieste, Italy; \email{marco.molinaro@inaf.it}}
\affil{$^2$Centre de Donn\'{e}es astronomiques de Strasbourg, Strasbourg, France}
\affil{$^3$Universität Heidelberg, Astronomisches Rechen-Institut, Heidelberg, Germany}
\affil{$^4$Friedrich-Alexander-Universität Erlangen-Nürnberg, Erlangen Centre for Astroparticle Physics, Erlangen, Germany}
\affil{$^5$University of Edinburgh, Edinburgh, Scotland, UK}
\affil{$^6$Centro de Astrobiología (INTA-CSIC), Madrid, Spain}

\paperauthor{Marco~Molinaro}{marco.molinaro@inaf.it}{0000-0001-5028-6041}{INAF}{Osservatorio Astronomico di Trieste}{Trieste}{}{}{Italy}
\paperauthor{Mark~Allen}{mark.allen@astro.unistra.fr}{}{CNRS}{CDS}{Strasbourg}{}{}{France}
\paperauthor{Fran\c{c}ois~Bonnarel}{francois.bonnarel@astro.unistra.fr}{}{CNRS}{CDS}{Strasbourg}{}{}{France}
\paperauthor{Fran\c{c}oise~Genova}{francoise.genova@astro.unistra.fr}{}{CNRS}{CDS}{Strasbourg}{}{}{France}
\paperauthor{Markus~Demleitner}{msdemlei@ari.uni-heidelberg.de}{}{Universit\"{a}t Heidelberg}{Astronomisches Rechen-Institut}{Heidelberg}{}{}{Germany}
\paperauthor{Kay~Graf}{kay.graf@fau.de}{}{Friedrich-Alexander-Universität Erlangen-Nürnberg}{Erlangen Centre for Astroparticle Physics}{Erwin-Rommel-Straße 1 Erlangen}{}{91058}{Germany}
\paperauthor{Dave~Morris}{dmr@roe.ac.uk}{}{University of Edinburgh}{}{Edinburgh}{}{}{UK}
\paperauthor{Enrique~Solano}{esm@cab.inta-csic.es>}{}{INTA}{CSIC}{Madrid}{}{}{Spain}
\paperauthor{Andr\'{e}~Schaff}{andre.schaaff@astro.unistra.fr}{}{CNRS}{CDS}{Strasbourg}{}{}{France}



\begin{abstract}
FAIR principles have \textit{the intent} to \textit{act as a guideline for those wishing to enhance 
the reusability of their data holdings} and \textit{put specific emphasis on enhancing the ability 
of machines to automatically find and use the data, in addition to supporting its reuse by individuals}.
Interoperability, one core of these principles, especially when dealing with automated systems’ 
ability to interface with each other, requires open standards to avoid restrictions that negatively 
impact the user's experience.
Open-ness of standards is best supported when the governance itself is open and includes a wide 
range of community participation.
In this contribution we report our experience with the FAIR principles, interoperable systems and 
open governance in astrophysics. 
We report on activities that have matured within the ESCAPE project with a focus on interfacing 
the EOSC architecture and Interoperability Framework.
\end{abstract}




\section{Introduction}

The Virtual Observatory (VO) ecosystem of standards and technologies, discussed and defined 
by the International Virtual Observatory Alliance (IVOA), has, since the setup of the Alliance 
itself (2002), focused on interoperability as a way to provide discovery of and access to distributed 
resources in the astrophysics research domain and to enable re-use of those same resources.

The above statement explains why VO standards provide a direct support of FAIR principles 
(Fig.~\ref{o4-001:ivoafair} provides a visual support on how FAIR principles fit in the IVOA 
architecture)
and, adding the fact that they're open in definition and governance, allow for an Open Science
scenario in astrophysics. VO also pre-dates the FAIR principles formalisation 
\citep{wilkinson2016fair} by a timespan long enough that let its architecture grow mature.

\articlefigure[width=.5\textwidth]{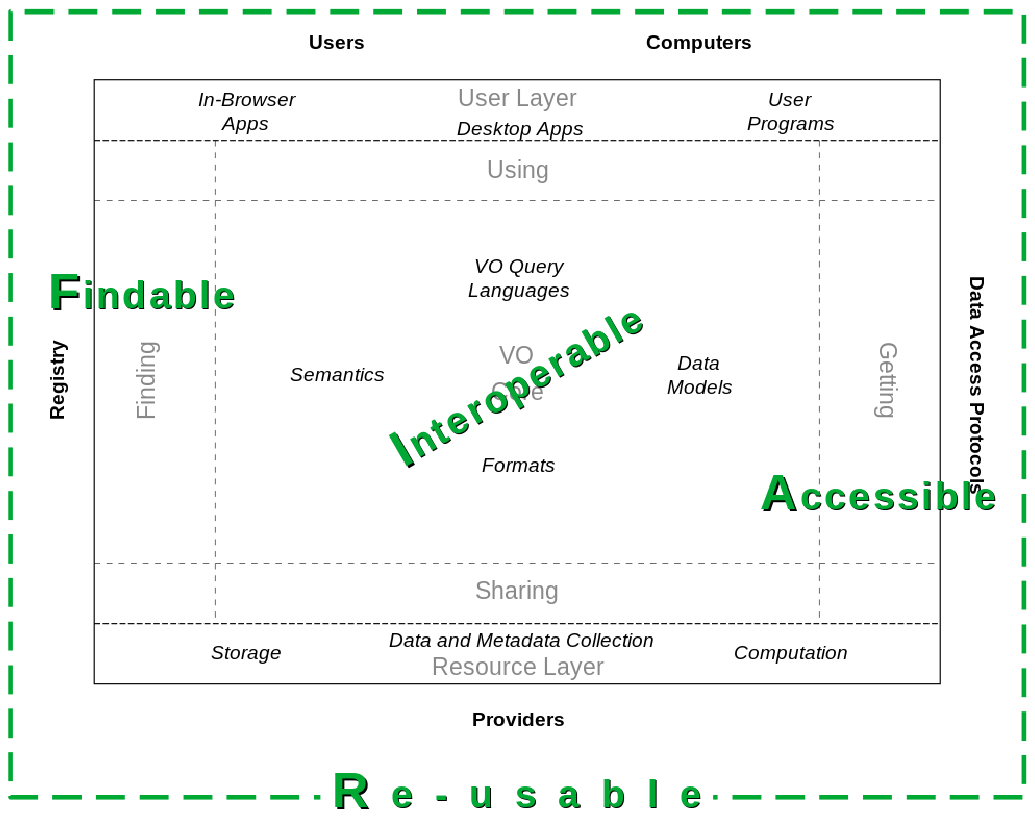}{o4-001:ivoafair}{The IVOA architecture (in black) 
and the top FAIR principles concepts superimposed (in green). Interoperability fits the core of the 
\textit{Level 1} IVOA Architecture \citep{2021ivoa.spec.1101D} while the pillars of Registry and Data 
Access Protocols embody the Findability and Accessibility. Re-usability pops out of the full 
architecture, taking advantage of the entire ecosystem the VO architecture defines.}

In Europe, the VO community coordinates itself around EuroVO, roughly spanning the same 
timeline of the gloabl IVOA efforts it contributes to. The succession of EuroVO projects 
\citep[sketched in Fig.~\ref{o4-001:eurovo}, adapted from][]{2015A&C....11..181G} is currently 
continued within the H2020 ESCAPE\footnote{EU Horizon 2020 funded: \url{https://projectescape.eu/}} 
project (see Sec.~\ref{sec:escape}), in particular within its Work Package 4 (WP4, CEVO 
``\textit{Connecting ESFRI projects to EOSC through the VO}'').

\articlefigure[width=.95\textwidth]{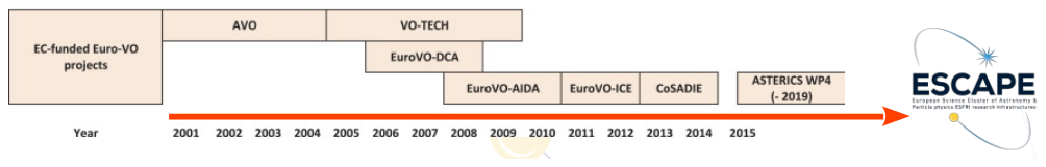}{o4-001:eurovo}{Timeline of EuroVO projects, 
adapted from \citet{2015A&C....11..181G}}

In this contribution the focus will be on the activities (Sections~\ref{sec:archint} to \ref{sec:software}) 
carried on within ESCAPE that deal with integrating the VO architecture in the European Open 
Science Cloud (EOSC) and supporting ESFRI projects in improving the interoperability of their 
resources, tackling common issues towards FAIR maturity.

\section{ESCAPE}
\label{sec:escape}

The ESCAPE project (European Science Cluster of Astronomy \& Particle Physics ESFRI 
Research Infrastructures) puts together astronomy and particle physics, through their 
Research Infrastructures considered crucial by the European Science Forum, towards 
the common goal of implementing a functional link between them and European 
Open Science Cloud (EOSC).
Within the project, the CEVO WP has the goal to connect ESFRI projects to EOSC using 
the VO framework as the means to do so. This involves three main acitivities: integration 
of astronomy VO data and services into the EOSC, implementation of FAIR principles for 
ESFRI data through the Virtual Observatory, and adding value to trusted content in 
astronomy archives.
The activities reported in the subsequent sections of this contribution are mainly 
performed within CEVO but include cross activities with other ESCAPE WPs.

\section{IVOA architecture integration}
\label{sec:archint}

\subsection{Registry integration}
\label{subsec:registry}

Integration of the IVOA Registry of Resources in EOSC has happened with the
EUDAT B2FIND service harvesting\footnote{Currently harvested records are visible at 
\url{http://b2find.eudat.eu/dataset?groups=ivoa}}, 
through the OAI-PMH protocol, DataCite metadata from the IVOA Registry. 
Work is ongoing towards adding more metadata, using the B2FIND metadata 
schema at harvesting, instead of DataCite only. The mapping from 
VOResource metadata to DataCite is not complete, the most important 
differences being about mapping per-protocol access URLs and tablesets.
The Unified Astronomy Thesaurus top level concepts 
have been mapped to VOResource subject keywords.
A further goal of such mapping and integration could be trying to bridge 
different disciplines of research. Initial use cases have been drawn for this 
purpose\footnote{You can find them here: \url{https://github.com/msdemlei/cross-discipline-discovery}}.

\subsection{EOSC onboarding}
\label{subsec:onboard}

The EOSC onboarding procedure shows another possible way 
of integration, with resources ending up directly visible in the EOSC portal.
To test this solution a Provider has been set up (for INAF) and a 
couple of services (in collaboration with the NEANIAS project) have 
been onboarded successfully. 
The onboarding procedure will be tested also for registered VO 
resources, to report guidelines on the procedure in view of
avoiding duplication of efforts for providers.
The differences in architecture, metadata goals and granularity
are the challenges offered to this activity.

\section{IVOA standards updates for ESFRI}
\label{sec:recupdate}

Interoperability and FAIR-ness inclusion in standards require
continuous updates.
In ESCAPE, CEVO continues gathering requirements from ESFRIs, using 
the IVOA interoperability meetings as milestones to report result and 
help keeping the IVOA standards up to date.
Radio astronomy data access, tools and standards to support follow-up 
of gravitational waves events are among the main results alongside 
common support to projects and communities.

\subsection{Science with interoperable data schools}
\label{subsec:schools}

Re-usability, besides its metadata implications, also requires a 
community of users that are aware of standards and tools.
 ESCAPE CEVO continues the sequence of Virtual Observatory schools 
 in Europe that has been an integral part of the EuroVO projects.
Gathering feedback from the community is a goal of these events 
alongside dissemination \citep[see][for details]{O5_001_adassxxxi}.

\section{Software interoperability}
\label{sec:software}

FAIR principles and interoperability are also useful guidelines for open 
solutions to discover, re-use and interoperate software and services.
Open-source scientific Software and Service Repository (OSSR, ESCAPE WP3)
activities include easing publication and integration of these non-data resources.
Work moves from code versioning on to subsequent publishing in dedicated 
repositories (like Zenodo and Docker Hub) and, finally, integration in the 
EOSC marketplace.
Software interoperability is also investigated in terms of standardisation 
of the current heterogeneous scenario of environments and tasks (Jupyter notebooks,
containers and other) that differ in terms of interfaces, behaviour, 
configuration, authentication.
Identifying a minimal set of common functionalities, trying to constrain 
the interfaces require test implementations to help in answering 
the first questions in interoperability of these environments\footnote{
For basic concepts see \url{https://wiki.ivoa.net/internal/IVOA/InterOpMay2021SPW/20210527-IVOA-canidothis.pdf}}.

\section{Connection to other projects, initiatives}
\label{sec:bridging}

Connections that ESCAPE members are making with other projects and initiatives 
help in keeping interoperability alive across domain boundaries (project, system or geographical).
Among these connections are worth mentioning that with the FAIRsFAIR project, with
contributions to activities in turning FAIR principles into reality, and with FREYA project, 
including discussion on Persistent identifiers usage starting from the experience of IVOA IDs.
Part of the activities have been also testing and discussion at RDA level about the FAIR Data
 Maturity Model and the proposed 
metrics connected, and the participation to LISA conference to report stewardship, curation and 
publication related to the ESFRI archives.

\acknowledgements ESCAPE - The European Science Cluster of Astronomy \& Particle Physics 
ESFRI Research Infrastructures has received funding from the European Union’s Horizon 2020 
research and innovation programme under the Grant Agreement n° 824064.
MM acknowledges support form the NEANIAS project in the EOSC onboarding procedure.

\bibliography{O4-001}  


\end{document}